# Atom Addition Reactions in Interstellar Ice Analogues

**H. Linnartz[a], S. Ioppolo[b], G. Fedoseev[a]**

[a] Sackler Laboratory for Astrophysics, Leiden Observatory, University of Leiden,

PO Box 9513, NL 2300 RA Leiden, The Netherlands

[b] Department of Physical Sciences, The Open University, Walton Hall,

Milton Keynes MK7 6AA, UK

**Abstract**

It was in 'The Magellanic Cloud' (1955) - a science fiction novel by Stanislaw Lem - that engineers traveling to another star noticed that their spacecraft for unknown reasons overheated. The cause had to be outside the spaceship, but obviously there was only emptiness, at least compared to terrestrial conditions. The space between the stars, the interstellar medium, however, is not completely empty and at the high speed of the spacecraft the cross-section with impacting particles, even from such a dilute environment, was found to be sufficient to cause an overheating. Today, 60 years later, the interstellar medium has been studied in detail by astronomical observations, reproduced in dedicated laboratory experiments and simulated by complex astrochemical models. The space between the stars is, indeed, far from empty; it comprises gas, dust and ice and the molecules detected so far are both small (diatomics) and large (long carbon chains, PAHs and fullerenes), stable and reactive (radicals, ions, and excited molecules) evidencing an exotic and fascinating chemistry, taking place at low densities, low temperatures and experiencing intense radiation fields. Astrochemists explain the observed chemical complexity in space – so far 185 different molecules (not including isotopologues) have been identified - as the cumulative outcome of reactions in the gas phase and on icy dust grains. Gas phase models explain the observed abundances of a substantial part of the observed species, but fail to explain the number densities for stable molecules, as simple as water, methanol or acetonitrile – one of the most promising precursor species for the simplest amino acid glycine - as well as larger compounds such as glycolaldehyde, dimethylether and ethylene glycol. Evidence has been found that these and other complex species, including organic ones, form on icy dust grains that act as catalytic sites for molecule formation. It is here where particles 'accrete, meet, and greet' (i.e., freeze out, diffuse and react) upon energetic and non-energetic processing, such as irradiation by vacuum UV light, interaction with impacting particles (atoms, electrons and cosmic rays) or heating.

This review paper summarizes the state-of-the-art in laboratory based interstellar ice chemistry. The focus is on atom addition reactions, illustrating how water, carbon dioxide and methanol can form in the solid state at astronomically relevant temperatures, and also the formation of more complex species such as hydroxylamine, an important prebiotic molecule, and glycolaldehyde, the smallest sugar, is discussed. These reactions are particularly relevant during the 'dark' ages of star and planet formation, i.e., when the role of UV light is restricted. A quantitative characterization of such processes is only possible through dedicated laboratory studies, i.e., under full control of a large set of parameters such as temperature, atom-flux, and ice morphology. The resulting numbers, physical and chemical constants, e.g., barrier heights, reaction rates and branching ratios, provide information on the molecular processes at work and are needed as input for astrochemical models, in order to bridge the timescales typical for a laboratory setting to those needed to understand the evolutionary stages of the interstellar medium. Details of the experiments as well as the astrochemical impact of the results are discussed.

## 1. Introduction

### 1.1 Ice in space

The interstellar medium (ISM) is filled with gas (99% in mass) and dust (1% in mass). The gas comprises hydrogen (H, 70%), helium (He, 28%), and a fraction of heavier elements; the dust mainly composes of silicate grains and carbonaceous material, remnants from evolved stars. The gas and dust are not distributed uniformly over the ISM. In translucent or diffuse clouds densities amount to a few 1000 particles/cm$^3$ and temperatures are as high as 100 K. With time, gravitation causes such clouds to collapse. New, denser clouds form, densities rise up to several hundred thousands particles/cm$^3$, the clouds turn opaque and as these regions are radiation shielded temperatures drop to 10-20 K. This is an important stage during the cosmochemical evolution of the ISM. Cores in the innermost regions of such dark clouds, ultimately, reach high enough densities and temperatures for a protostar, i.e., a future star, to be formed inside. Conservation of angular momentum, subsequently, causes the formation of a so called protoplanetary disk around the forming star with perpendicular outflows. At this stage, the star has not been born yet, but the cloud character has been lost, reason why this is called a 'young stellar object' (YSO). For solar mass stars it takes about 10 millions years before a star finally accretes most of the nearby material and reaches its final mass, while the left-overs in the protoplanatery disk provide the material from which planets and other celestial objects form.

This review paper mostly focusses on laboratory studies of chemical processes during the very first steps of star formation, i.e., the dark cloud stage and the formation of YSOs. This focus has two reasons: first, microwave, submillimeter and infrared (IR) observations toward dark clouds show that these regions already possess a rich chemical inventory even before stars and planets are formed. Second, an exact understanding of the chemical processes involved in the formation of interstellar species observed in these regions offers a tool to link the chemical processes in space to the chemical composition of (exo)planets. Ultimately, this connects the presence of organic compounds to the emergence of life. The solid state plays an important role in this evolutionary stage, as gas phase processes alone cannot explain the observed abundances of many of the identified species; important astrochemical processes must take place on icy dust grains.

At the low temperatures in dark interstellar clouds, dust grains act as (sub)micrometer-sized cryopumps onto which gas-phase particles accrete. In dense cores - the innermost part of molecular clouds - the approximate timescale at which gas-phase species deplete onto grains is $10^5$ years, comparable to the lifetime of a dark cloud. Therefore, during the first stage of star formation, most of the molecular species, with the exception of the lightest ones ($H_2$ and He) are expected to be largely frozen-out onto interstellar grains, as sticking probabilities are close to unity. The resulting ice mantle provides a local molecule reservoir, which may be as thick as several tens of monolayers. Moreover, the icy dust grain provides a 'third body' to which two reactive species can donate excess energy allowing formation of a stable molecule, a process that otherwise is prohibited in a two body gas-phase collision. Although accretion rates on the surface of interstellar grains are extremely low, astronomical timescales allow grains to accumulate enough material to become chemically active.

Excellent text books and reviews exist that describe in much detail the physics and chemistry of the processes taking place in the interstellar medium, both in the gas phase and in the solid state (see e.g., [001-005]).

The interstellar ice composition has been derived from mid-infrared observations of molecular clouds along the line of sight to a background star. Also a newly formed protostar embedded into a young stellar object can be used as a light source. The ice absorption spectra are obtained by subtracting the emission profile of the background star or embedded protostar from the observed spectra (see e.g., [006-008]). The features leftover correspond to the absorbance of molecules in the line of sight between the emitting object and the observer. Observations in the mid-IR range from ground-based telescopes are limited to a few spectral windows, because of the presence of telluric lines in the Earth atmosphere. This limitation is overcome by airborne and space observations. Observations with the Infrared Space Observatory (ISO) [006,009,010] and the Spitzer Space Telescope [007,008,011,012] greatly improved our knowledge on interstellar ice composition by exploring wavelength ranges inaccessible from Earth. The interpretation of these data only has become possible following ice data recorded for different settings in the laboratory (see e.g., [013-017]). The systematic laboratory investigation of infrared band profiles – peak position, band width and integrated band ratios - of pure and mixed ices (i.e., containing two or more species) over a wide range of temperatures combined with knowledge of the corresponding absorption band strengths allows for the identification of the components of the interstellar ice mantles.

The main components of interstellar ices are $H_2O$, CO, $CO_2$, $CH_3OH$, $NH_3$, $CH_4$, XCN (reflecting components with a CN stretch), and likely HCOOH and OCS. These ices have generally an amorphous, compact structure and spectroscopic detections connect solid state species that are chemically linked, such as CO and $CH_3OH$. The spectra also reflect sequential accretion events that result in layered structures, with the most volatile species accreting last. The first layer on top of the grain is a $H_2O$-rich polar-ice with embedded $CO_2$, $CH_4$, and $NH_3$. The observed abundances of these species correlate with the abundance of solid $H_2O$. The second (top)layer is a non-polar CO-rich ice [008,016,018-020]. Spectroscopic evidence has been found hinting for the presence of more complex molecules, like HCOOH, $CH_3CHO$, and $C_2H_5OH$ [007,008,021,022]. However, better constraints, i.e., higher quality spectra, especially at 5-10 μm, are needed to unambiguously identify these species in the solid state. Moreover, mid-IR observations are insensitive to homonuclear diatomic molecules, like $O_2$ and $N_2$. Molecular oxygen is actually not expected to be present in the ice due to its chemical reactivity, but nitrogen may be in fact one of the more abundant ice components, because of its chemical inertness and stability.

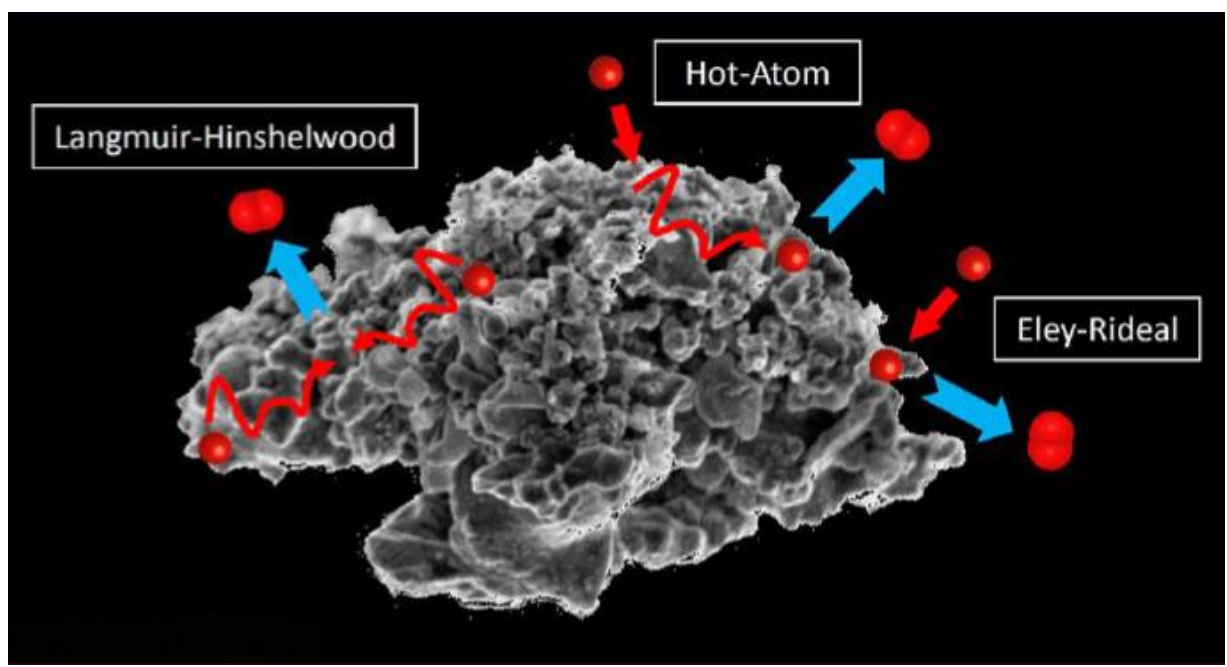

**FIGURE 1 –** *A schematic of the chemical processes taking place on the icy surface of a cold interstellar dust grain.*

### 1.2 Ice processing

Chemical reactions in and on interstellar ices take place through neutral-neutral atom and radical addition reactions following the accretion of species, or as a result of energetic processing (by impacting photons, cosmic rays, free electrons, etc.). The first group is interesting from an astronomical point of view, since solid-state radical-radical and radical-molecule reactions often do not require any activation energy; they can occur even at extremely low temperatures (10 K), that is, in the innermost parts of molecular clouds, where newly formed molecules are largely shielded from radiation by dust particles. As such they provide the chemical starting box for processes to take place at a later stage in the star formation process, when also UV fields increase in intensity. A systematic overview of these different processes is provided below.

<u>Atom and radical addition reactions.</u> Among the possible atom addition reactions (e.g., H, O, N, C, **S,** F, Cl etc.), the most important one is the hydrogenation of ice. This is not surprising since hydrogen is the most abundant element in the Universe. Some of the species formed through hydrogenation reactions on the surface of dust grains are $H_2$ (through H+H), $H_2O$ (through hydrogenation of $O/O_2/O_3$), $NH_3$ (through hydrogenation of N atoms), $CH_3OH$ (through hydrogenation of CO), and possibly $CH_4$ (through the hydrogenation of C atoms). $CO_2$ can be formed through atom and radical addition reactions: CO+O and CO+OH.

Reactions taking place on the surface of the grains are believed to proceed through one of three possible mechanisms. These are Langmuir-Hinshelwood (L-H), Eley-Rideal (E-R), and "hot-atom" (or Harris-Kasemo) mechanisms [023] (see Figure 1). In the case of the L-H mechanism, two reactants are adsorbed on different surface spots of an interstellar grain and reach thermal equilibrium with the surface before one or both diffuse to find each other and react. In the E-R case, one of the reactive species directly lands on top of the other and the reaction immediately occurs before thermal equilibrium is reached. The hot-atom mechanism is a combination of both: one species is located on the surface in thermal equilibrium, while the second one lands on a nearby surface spot; migration and reaction take place before thermal equilibrium is reached.

Distinguishing between Langmuir-Hinshelwood, Eley-Rideal, and hot-atom mechanisms is of great importance when the temperature of the adsorbing species is significantly higher than the temperature of the surface. For the E-R and the hot-atom mechanisms, the high energy of the incoming species helps to overcome the activation barrier of the reaction. This is not true in the case of an L-H event, where both reactants are in thermal equilibrium with the ice surface. In cold dark molecular clouds, the difference in temperature between the accreting species and the ice surface is not large. Because of the extremely low accretion rates even in the densest regions of molecular clouds, the L-H mechanism is the most important one in non-energetic surface reactions. However, for laboratory conditions, the temperature of the exposing atom beams is often equally high or even higher than room temperature. This deviation from conditions typical in space has to be taken into account during the interpretation of laboratory data. Subsequently, Monte Carlo (MC) simulations [024] can be used to simulate and reproduce laboratory results to investigate qualitatively and quantitatively all the mechanisms at play in the solid state (e.g., deposition, diffusion, segregation, reaction, and desorption). Results from these models can then be included into astrochemical models to extend laboratory results to interstellar timescales [025-027].

The experimental investigation of non-energetic reaction routes as discussed in this paragraph is the main subject of this review paper. Many of the results presented here are based on work described in [028,029]. The main ice reaction routes with energetic input are shortly addressed below for completeness and representative references are given .

UV-induced reaction routes. As opposed to the neutral-neutral, atom, and radical addition surface reactions discussed above, UV processing of ices provides an energetic input that leads to the excitation of molecules and dissociation of molecular bonds followed by the formation of "hot" fragments, i.e., radicals, atoms, and ions with high internal energy. These "hot" fragments can in turn react with the surrounding molecules and easily overcome activation barriers, photodesorb from the ice surface, or diffuse into the bulk of the ice. The very first experiments showed that UV-photon induced ice chemistry can efficiently lead to the formation of complex organic species in the ice mantles. At high doses and high enough ice thicknesses, using ices made up of a number of different constituents, a complex polymeric refractory residue was found to form containing biologically relevant species, including amino acids [030-034]. Such 'top-down' experiments were initially taken as a proof that molecular complexity in space forms in the solid state, but did not allow investigating the individual processes at play. More recent 'bottom-up' approaches, investigating reaction pathways selectively in pure or binary ices, make it possible to study UV induced processes at a higher level of accuracy [035]. Photodesorption is considered to be an important non-thermal sublimation process that explains the presence of gas-phase species in the ISM at temperatures far below their accretion value. Photodesorption rates for a number of molecules - CO, $H_2O/D_2O$, $CO_2$, $N_2$, $O_2$, $CH_3OH$, and several more - have been determined both experimentally and theoretically [036-048] and the underlying processes have been characterized by studying the wavelength dependency using tunable VUV light sources [049-052]. Although there exist some discrepancies between values derived by different groups, likely due to different spectral emission patterns of the used light sources, the overall message is that the photo-induced desorption efficiency is substantially higher than assumed in many of the initial astrochemical models. Other studies have been focusing on the photo-induced reactions taking place in the ice. Extensive studies [039,053,054] show how Ly-alpha irradiation of pure methanol ice results in the formation of larger complex species, including glycolaldehyde and ethylene glycol, and how water ice affects the reaction efficiencies. As stated before, methanol is effectively formed upon CO hydrogenation, therefore, methanol is expected to be mixed with CO in the ice in the ISM [016].

So far, many studies have been performed for photons with an energy around 10.2 eV (121.6 nm), using microwave driven $H_2$ discharge lamps. In the ISM, radiation fields can indeed be Ly-alpha dominated, during some evolutionary stages, via cosmic rays induced excitation of hydrogen emitting such light. A detailed analysis of the UV-induced reaction network is often challenging without the knowledge of photodissociation and photoionisation cross-sections for all the molecules involved in the UV photolysis process. Furthermore, these cross-sections and photodissociation branching ratios can be wavelength dependent. Other processes, like structural changes in the ice morphology also have to be taken into account, but this falls outside the topic of this review. Important to note is that UV photolysis influences the chemical evolution of the ISM by inducing higher molecular diversity in the solid phase and allowing at the same time desorption of molecules even in cold dense regions, i.e., environments where such species are expected only in the solid phase.

<u>Cosmic rays.</u> The chemical composition of the ices can be further modified by energetic particles, i.e., through highly charged ion and electron bombardment. The energy released by a highly energetic ion passing through a material causes the dissociation of hundreds of molecules along its path. These fragments can then recombine. As a result of the interaction between a cosmic ray and the ice, many new molecules including both simple and more complex species are formed. Moreover, the impact of a cosmic ray with the ice surface causes the release of (some) ice material into the gas phase through sputtering and local heating. In different laboratories, the effect of such impacts has been extensively studied [055-061]; recent work comprises [053, 062-067]. Obviously, the impact of a cosmic ray impact is larger than that of a UV-photon or H-atom. However, the flux of UV photons in the ISM is substantially higher than that of cosmic rays and a direct comparison is only possible when also taking into account the relative values typical for molecular clouds.

Thermal processing. During their evolution, star-forming regions undergo thermal processing. Most of the chemistry in interstellar ices takes place within the range of temperatures 10-100 K. These are cryogenic temperatures; however, due to the very long time scales involved, thermally induced chemistry still may greatly affect the composition of the ices. Thermal processes can induce sublimation of ice material [068-070], changes in the ice morphology [071], segregation of species in the ice [072], neutral-neutral reactions, such as deuterium-exchange reactions [073,074], proton transfer reactions [075-077] and thermally induced nuclefilic reactions [078,079].

All aforementioned mechanisms dominate in different regions of the ISM or at different evolutionary stages in star-forming regions, but all mechanisms have in common that they contribute to a higher molecular complexity in space. The solid state (surface) formation of $H_2$ is very important and a topic on its own. It will not be further discussed here and we refer to [080] that provides a good review. The emphasis in this review will be on other species formed in solid state (ice) atom addition reactions.

The next section focuses on laboratory work, simulating atom-addition reactions, first with an introduction on the used experimental concepts, summarizing the state-of-the-art, and second, with details of an experimental setup fully devoted to the study of atom addition reactions in ice.

## 2 Experimental

## 2.1 Measurement concepts

Current state-of-the-art experiments on interstellar ice analogues use ultra-high vacuum (UHV $\sim 10^{-10}$ mbar or lower) surface techniques. Under such vacuum conditions, build-up of ice contaminations from background gas is slow (<1 monolayer (ML) per hour). Consequently, deposited ices are minimally contaminated and this allows studying ices with monolayer precision. Ices are typically grown on a cryogenically cooled surface with the substrate holder mounted on the tip of the cold head of a closed-cycle He cryostat. The minimum temperature depends on the employed cryostat, as well as on the details of the substrate and substrate holder. Most contemporary experiments have minimum temperatures of $\sim 10$ K, a value on the lower side for interstellar ice analogues. The temperature is controlled through resistive heating, with access to values as high as several hundreds of K.

*In situ* deposition of ices through condensation of volatiles, admitted into the vacuum chamber at specific pressures and for a specific period of time, enables control of ice thickness, ice mixing ratio and porosity. Experiments have shown that the ice structure strongly depends on the deposition angle, with more compact ices forming from direct deposition onto the surface and more porous ices forming from depositions from other directions. A special case is diffusive deposition, where the molecular beam is directed away from the substrate to promote depositions at random angles, generally further increasing the porosity level. The final ice thickness can be determined by laser interference patterns, while ice morphology can be approximated from the experimental conditions and by means of mass spectrometric techniques (exploiting the fact that trapping efficiencies of volatiles in less volatile ices increase with porosity) or infrared spectroscopic techniques (by comparing ice surface and ice bulk vibration modes, e.g., the strength of the dangling OH feature in $H_2O$ ice).  For a recent overview see [081].

The nature of the substrate varies between different experiments, depending on the object of the study and the employed analysis methods. Silicate and graphite surfaces most closely resemble the expected composition of interstellar grains, but have lower transmission and reflectance compared to many other materials, which limits diagnostic methods. Both surfaces are also chemically active, which complicates the analysis. Gold and other noble metals are chemically inert and highly reflective, enabling techniques based on reflection spectroscopy. Any processes that depend on the characteristics of the metal surface itself (e.g., the release of secondary electrons during UV irradiation) are obviously not relevant to interstellar environments and should be considered carefully. Transparent windows constitute a third family of common substrates because of the possibility of transmission spectroscopy. The surface normally matters only for very thin ices, as here the interaction between surface and ice is directly involved. For typical interstellar ices comprising tens of monolayers the influence of the substrate is generally neglected.

The methods that are used to investigate reactions in interstellar ice analogues can be roughly divided in two groups, depending on analysis technique and thickness regime (submonolayer vs. multilayer). In the monolayer regime, the reactants and products can be probed mass spectrometrically. The surface is initially exposed to a small quantity of reactions, after which a temperature-programmed desorption (TPD) experiment is performed. In such an experiment, an ice sample is heated linearly, typically with a rate of 0.1 to 10 K / min [068]. The desorbing species are recorded as a function of temperature using quadrupole mass spectrometry (QMS). Alternatively laser photoinduced desorption can be used instead of a TPD. In this case, formed species can be quantified at the desired sample temperature and kinetic curves are obtained [082,083]. Different initial exposures and temperatures are probed to obtain information on, for instance, the reaction order. The TPD technique is very sensitive and permits submonolayer exposure. Furthermore, it allows the study of reaction products that do not remain on the surface or that do not form an ice, like the formation of $H_2$ from H atoms. The method however suffers from a few disadvantages: the products cannot be probed *in situ*, additional reactions during the heat-up to desorption cannot be excluded, quantifying the desorbing species is not straightforward, some of the species have equal, i.e., undistinguishable masses, and a TPD experiment is only possible through evaporation, i.e., destruction of the ice. The second method is to grow an ice of several monolayers and to expose this ice while recording reflection absorption infrared spectra (RAIRS). For this the infrared beam is guided in reflection mode through the ice, passing it twice. In this way, the reactants and products are probed *in situ* at the time and temperature of interest, which is the main advantage of this technique. Quantifying the formed product is only straight forward, provided that the RAIRS is calibrated with an independent method. In transmission mode, infrared band intensities of ices can be calibrated using laser interference techniques to derive absorption coefficients that can subsequently be used together to derive ice thicknesses in a range of experiments. These absorption coefficients are tabulated for many common ices in literature. A major complication is that many ice bands are sensitive to the local environment and an absorption coefficient derived for a pure crystalline ice is not *a priori* accurate for mixed ices or even amorphous and/or porous pure ices. When this technique is used to derive absolute ice abundances, great care must be taken to calibrate it for different kinds of ices. If well calibrated, spectroscopic techniques are very powerful tools, since they provide detailed information about the changes of ice composition and restructuring during irradiation and thus can be used to probe all kinds of ice processes as they occur, obtaining detailed kinetic information. The main disadvantages are that not all species can be detected in this way (e.g., when a species lacks a dipole moment or when bands overlap) and that the sensitivity is inferior to the TPD technique. The RAIRS detection technique is therefore often complemented by TPD, as it is possible to perform both methods simultaneously. A good example of this approach is available from [039]. A further extension of the method is to use isotopically labelled precursors – D, $^{13}$C, $^{15}$N or $^{18}$O. This allows recording additional spectroscopic and mass spectrometric data and moreover, excludes that reaction products are due to background gas pollutions (see also [023]).

### 2.2 State-of-the-art

In 1982, Tielens and Hagen [084] introduced an extended gas-grain model in which water ice – the dominant solid state species - is produced by sequential hydrogenation (i.e., H-atom addition) of O-containing ices (O, $O_2$, and $O_3$). This first stage of grain-surface chemistry results in the formation of a polar (water-rich) ice. In the next phase, along the gravitational collapse, the density in the molecular cloud increases, and CO freezes-out on top of the grains, forming an apolar (water-poor) ice layer on top of the polar one. Subsequent hydrogenation leads to the formation of formaldehyde and methanol. The existing astrochemical surface reaction networks (see e.g., Wakelam et al. [004] and references therein) were largely based on chemical intuition and gas-phase equivalent reaction paths. This changed with the introduction of cryogenic UHV techniques in solid state astrochemistry, around the start of this century. Over the last 15 years, particularly five specific surface reaction schemes have been studied in much detail: *i)* the formation of hydrogen peroxide and water upon hydrogenation (deuteration) of oxygen-containing ice, i.e., O+H, $O_2$+H and $O_3$+H [085-102]; *ii)* the formation of formaldehyde and methanol upon sequential hydrogenation of CO ice [025,103-114]; *iii)* the formation of carbon dioxide, mainly following CO+OH and CO+O [115-125]; *iv*) the formation of hydroxylamine NH$_2$OH [126-129]; *v*) and very recently the ongoing hydrogenation of CO producing glycolaldehyde and ethylene glycol [130]. Other studies report the hydrogenation and oxidation/nitrogenation of various nitrogen oxides such as NO and NO$_2$ [131-134]; formation of OCS *via* interaction of CS$_2$ with oxygen atoms [135]; the reactivity between non-energetic hydroxyl (OH) radicals and methane (CH$_4$) [136]; the investigation of the reactions of H and D atoms with solid C$_2$H$_2$, C$_2$H$_4$, and C$_2$H$_6$ [137]; the formation of ammonia (NH$_3$) [138-141], isocyanic acid (HNCO) [140,142], formic acid (HCOOH) [143], carbonic acid (H$_2$CO$_3$) [144], ethanol (CH$_3$CH$_2$OH) [145]; and oxygen atom reactions with alkenes [146]. In this review, a number of proto-typical examples will be discussed in more detail to illustrate representative experiments and to discuss their astrochemical relevance.

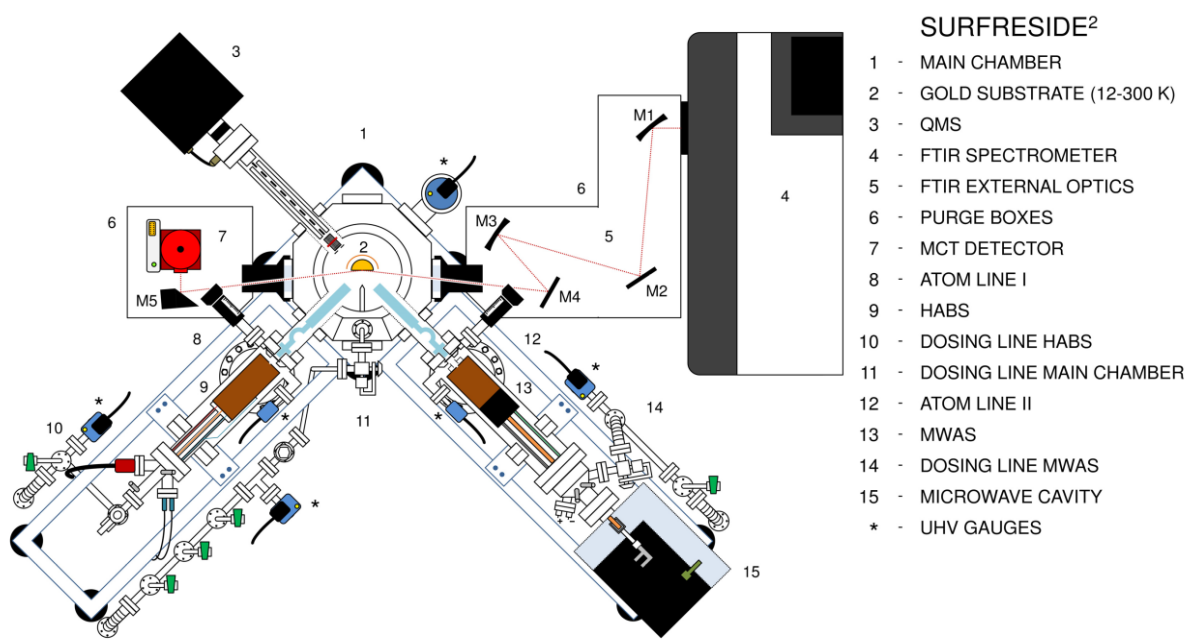

**FIGURE 2** – *Schematic top-view of SURFRESIDE$^2$ (Reproduced with permission from [125] - Copyright [2013], AIP Publishing LLC).*

### 2.3 SURFRESIDE[2]

Several setups worldwide are capable of studying atom- (and radical) addition reactions in interstellar ice analogues under ultra-high vacuum conditions [095,122,139,146-149]. These setups typically comprise one or two atom beam lines suited to study hydrogenation, and simultaneous or sequential atom (H/D, O, N) addition reactions. The principle procedures used in these setups are rather similar and largely based on RAIRS and TPD, as discussed above. Here, SURFRESIDE[2] is described, a recently extended setup in Leiden (NL) that in many respects is representative for the methods and techniques used in other laboratories. Full details are available from [028,029,125,141,150].

The central part of the SURFace REaction Simulation Device (Figure 2) is an UHV chamber ($10^{-10}$-$10^{-11}$ mbar) onto which two atom beam lines are mounted that aim at an optically flat gold coated cupper substrate (2.5 x 2.5 cm$^2$) placed in the centre of the main chamber, and mounted on the tip of a cold-head with temperatures accessible between 12 and 300 K using resistive heating. Ices are grown using two all metal high-vacuum stainless steel dosing lines, allowing different gasses to be deposited separately or simultaneously onto the gold substrate. This surface is not representative for an interstellar dust grain, but offers a chemically inert and effective heat conductor that in the infrared is highly reflective. Moreover (see section 2.1), ice thicknesses in space and in the laboratory are typically tens of ML thick, which substantially reduces the role of the substrate onto which the ice has been grown. Indeed, the use of other surfaces (such as HOPG) does not show different processes as long as these take place in the upper layers of the bulk ice. The deposition proceeds with a controllable flow that allows growing ices with monolayer precision.

The ices are monitored both spectroscopically by RAIRS using the light of a Fourier transform infrared spectrometer (700-4000 cm$^{-1}$, 0.5-4 cm$^{-1}$ resolution) and mass spectrometrically by TPD using a quadrupole mass spectrometer that is mounted behind the substrate and opposite to one of the two atom beam lines. The FTIR path is purged by dry air to minimize atmospheric pollutions. Structurally different experiments can be performed with a system like SURFRESIDE$^2$. During *pre-deposition*, ices are first deposited and subsequently exposed to atoms or radicals produced by the atom beam lines. Newly formed species in the ice are monitored with respect to a reference spectrum of the initially deposited ice, using RAIR difference spectroscopy. During *co-deposition*, molecules and atoms/radicals are deposited simultaneously onto the substrate. By changing the molecule/atom ratio different reaction stages in the selected chemical network can be simulated. In this case, RAIR difference spectra are obtained with respect to a bare gold substrate spectrum. After (co)deposition, a TPD can be performed by linearly heating the sample until the ice is fully desorbed. The thermal desorption is monitored using RAIRS. Alternatively, when the sample is rotated to face the QMS, the latter can be used to monitor the evaporating species, and to constrain the RAIRS data.

SURFRESIDE$^2$ uses two different and commercially available atom sources: a thermal cracking source (HABS - Hydrogen Atom Beam Source [151]), and a microwave discharge atom source (MWAS - MicroWave Atom Source [152]). The first one generates H- and D-atoms, while the second one can also produce N- and O-atoms and radicals such as OH. Before impacting onto the ice, the atoms are first cooled to room temperature into a nose-shaped quartz pipe that is designed to quench excited species by collisions with its walls. In this way 'hot' species do not reach the ice. However, as H collisions cause atoms to recombine to H$_2$, and as a consequence it is far from trivial to determine the impacting fluxes accurately. The relative high remaining temperature of 300 K does not affect the reaction efficiency in the case of a Langmuir-Hinshelwood mechanism, as species are believed to thermalize after hitting the ice surface, i.e., before they can diffuse and react. Calibration methods used so far are based on determining absolute atom densities in the exposing atom beam or, alternatively, on measuring the formation yield of final products in barrierless surface reactions followed by evaluation of original atom flux. Both these methods have been described in detail [125]. Fluxes can be varied by changing the pressure and power settings. Typical fluxes amount to $10^{12}$-$10^{13}$ H-atoms cm$^{-1}$s$^{-1}$ (HABS) and $10^{10}$-$10^{12}$ atoms N- and O-atoms cm$^{-1}$s$^{-1}$ (MWAS). Absolute flux values have errors of the order of 30-50 %. These uncertainties are large, but absolute flux values are needed to interpret reaction rates, quantitatively. The relative errors are generally much lower (~10%) and this still allows to compare deposition ratios.

It should be noted that in a pre-deposition experiment, an ice is bombarded after deposition. This generally limits reactions to the surface layers depending on penetration depths in the ice. In a co-deposition experiment, ice molecules and atoms or radicals are deposited simultaneously and depending on the applied ratio this allows each deposited species to be available for the reaction. The latter is generally closer to the processes taking place in space (although not exclusively) and as previously mentioned often allows to monitor both intermediate and final products of the reactions [093,102,128,130].

Below, a number of specific examples are discussed. These are largely biased by recent and ongoing work at the Sackler Laboratory for Astrophysics in Leiden, the Netherlands (www.laboratory-astrophysics.eu), but we have tried to put these results within the context of work ongoing in other groups active in the same field.

### 3. Results

### 3.1 O/O₂/O₃+H

Water is the main constituent of inter- and circumstellar ices and is an essential ingredient for life on Earth. In dense cold regions of the interstellar medium, gas-phase formation routes for $H_2O$ and subsequent freeze-out mechanisms cannot explain the large ice abundances observed. It is therefore expected that water forms in the solid state [153]. This idea was first proposed by van de Hulst [154], who indicated that water molecules are formed on the surface of grains by hydrogenation of $O_2$ molecules. Decades later, Tielens and Hagen [084] suggested that water forms upon sequential hydrogenation of atomic oxygen, molecular oxygen and ozone on icy dust surfaces. At that time, only gas-phase data were available, and this severely affected the completeness of the solid state reaction scheme and its correct incorporation into astrochemical models. It is only since a few years, starting in 2008, that the first fully controlled laboratory experiments on surface hydrogenation of oxygen precursors were actually performed. Miyauchi et al. [086] presented results for $O_2$+H at 10 K, while the work of Ioppolo et al. [087] comprises a large set of measurements covering 12 to 28 K. In both studies – bombardment of a pre-deposited $O_2$ ice with H- (D-)atoms – the formation of water ice via hydrogen peroxide was proven effective by RAIR techniques that permitted to derive reaction rates also for the deuterated species. Furthermore, Ioppolo et al. [087] showed that these rates were surprisingly independent of the studied temperatures.

In a follow-up study [092,093] it was confirmed that the $H_2O_2$ and $H_2O$ formation rates are temperature and thickness independent, whereas the final total yield increases with increasing temperature and thickness. This can be explained by a competition between reaction of H atoms with $O_2$ molecules (temperature independent) and hydrogen diffusion into the ice (temperature dependent): H atoms penetrate into the bulk of the ice with a penetration depth that is temperature dependent, and, once trapped in the ice, reaction is likely to occur. Consequently, the $O_2$ ice is mainly hydrogenated bottom-up.

Ioppolo and coworkers [092] also concluded that the reaction scheme initially proposed by Tielens and Hagen [084] could not be fully complete, as clear evidence was found in the RAIR spectra for $O_3$ formation upon hydrogenation of pure $O_2$ ice. This observation was explained by Cuppen et al. [093], who performed co-deposition experiments in which the $H/O_2$ ratio was varied in order to trace each step of the reaction scheme, selectively. This experiment is illustrated in Figure 3.

The ratio between the deposited H atoms and $O_2$ molecules determines the hydrogenation grade. Four H atoms are needed to convert $O_2$ into two $H_2O$ molecules. The four panels show decreasing $H/O_2$ ratios from top (100) to bottom (1). The top spectrum is clearly dominated by broad water and hydrogen peroxide bands, i.e., the final products of the reaction scheme as expected in a fully hydrogen dominated regime. In the oxygen dominated regime (bottom spectrum), i.e., the domain where full hydrogenation cannot be reached anymore by a continuous supply of $O_2$, intermediate species – $HO_2$, $H_2O_2$, and OH – become visible. The first intermediate ($HO_2$) is clearly visible in the $H/O_2=1$ experiment. Reducing the $O_2$ flow by a factor of two causes the $HO_2$ signal to decrease, whereas $H_2O_2$ and OH signals appear to increase somewhat. A further reduction ($H/O_2=10$) results in broad bulk water bands and the intermediate species are not visible anymore. A more complex and complete reaction scheme was derived by tracing the abundance of each single species involved in the reaction scheme versus the H-atom fluence for different temperatures and $H/O_2$ ratios. This way, Cuppen et al. [093] could explain, for instance, the presence of ozone in the ice through the formation of atomic oxygen by the reactions $HO_2+H \rightarrow H_2O+O$ that was not previously included in the scheme, and the subsequent reaction $O+O_2 \rightarrow O_3$.

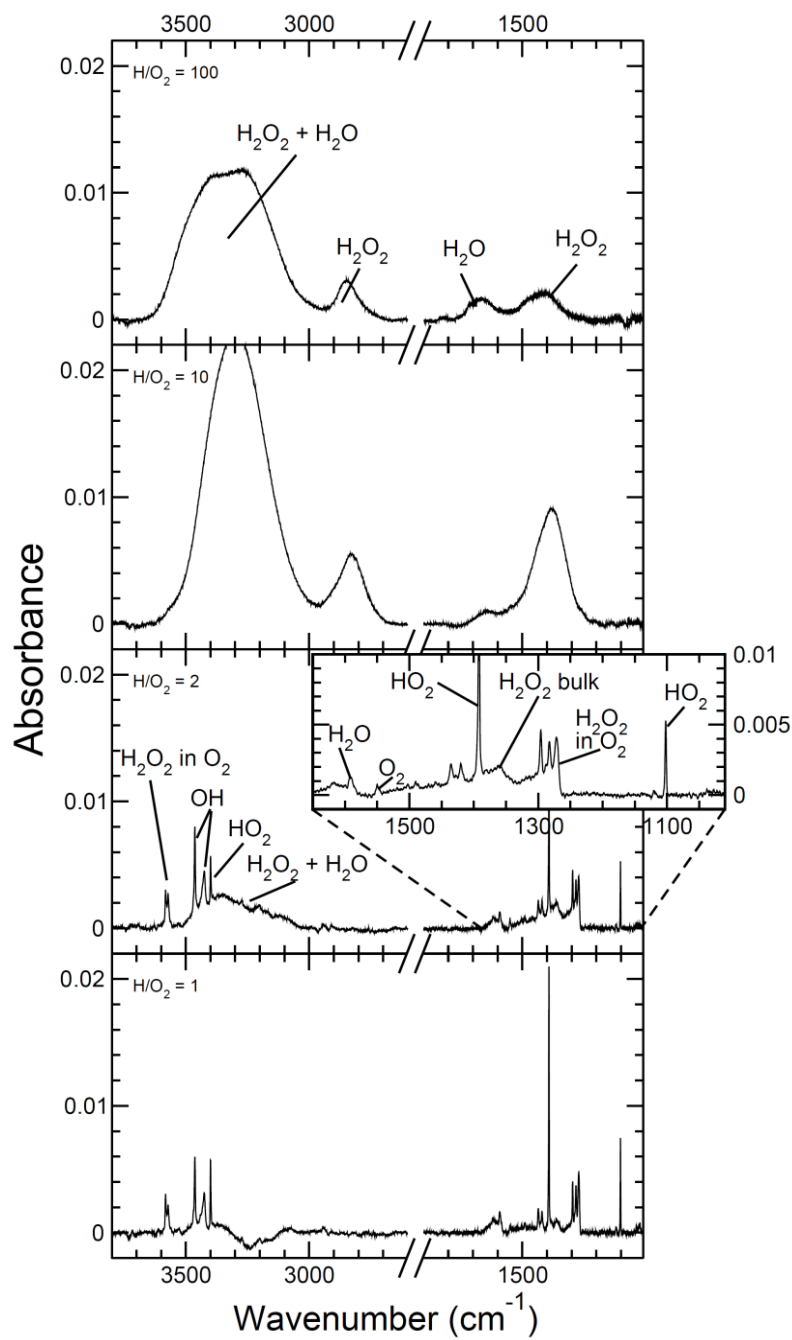

**FIGURE 3** – *RAIR spectra of H and $O_2$ co-deposition experiments performed at 20 K and for different $H/O_2$ ratios of 100, 10, 2, and 1 from top to bottom. The H-atom flux is 2.5 x 10$^{13}$ atoms cm$^{-2}$ s$^{-1}$, and is kept the same for all the experiments. (Figure taken from [093] - reproduced by permission of the PCCP Owner Societies).*

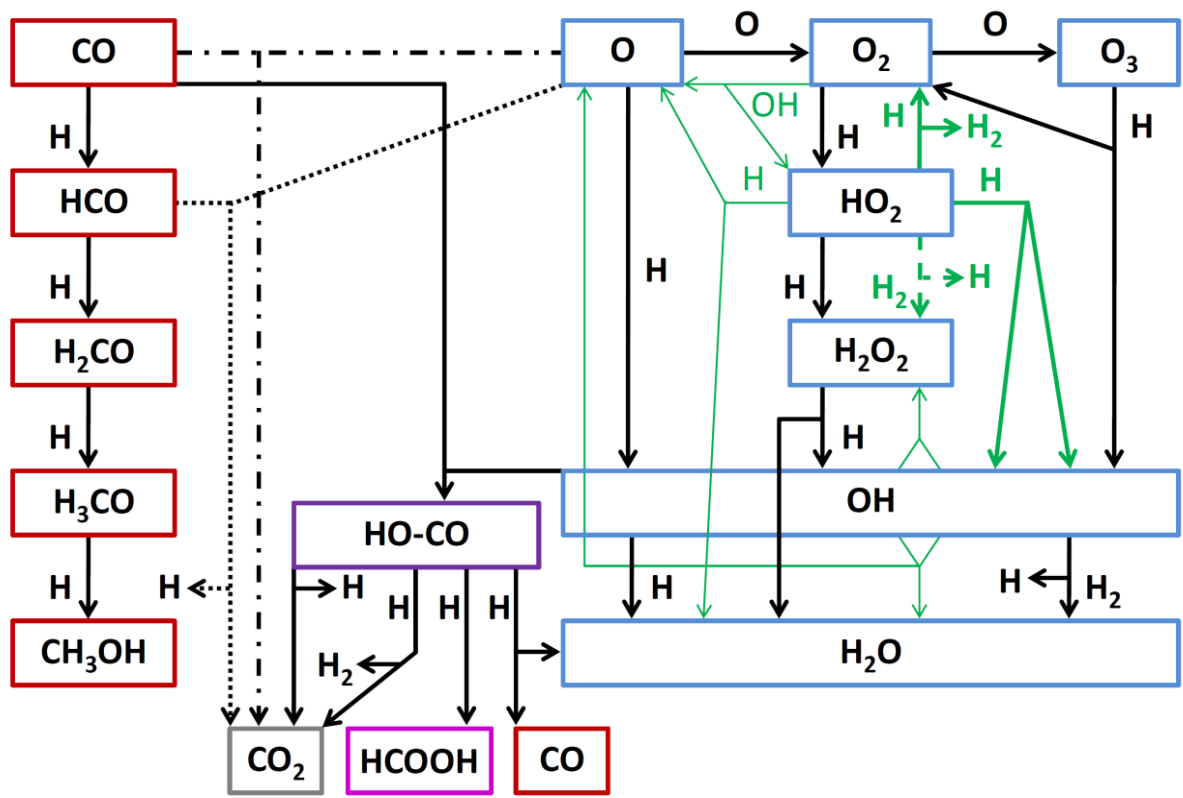

**FIGURE 4** – *On the right side a schematic representation is shown of the reaction network involving $O/O_2/O_3$ hydrogenation as discussed in the text and shown in Figure 2 for different co-deposition experiments. Other reactions – that are discussed later - are shown as well. On the left side the primary steps in the hydrogenation of CO are shown to result in the formation of formaldehyde ($H_2CO$) and methanol ($CH_3OH$) (Section 3.2). When bombarding a mixed ice, comprising both CO and $O_2$, also $CO_2$ is formed (middle panel) through de-hydrogenation of the HO-CO reactive intermediate (Section 3.3). From this complex other species, including formic acid, can be formed (Section 3.4). Not shown in this figure is that the ongoing hydrogenation of CO also can result in more complex molecules (Section 3.5). The diagram illustrates that the processes at play – even in a binary mixture - are already quite complex. The figure is adapted from Ioppolo et al. [120].*

The resulting reaction channels derived from all these experiments are indicated in Figure 4. It extends the diagram initially proposed by Tielens and Hagen [084] with a number of extra channels, including a cross channel between $O_2$ hydrogenation and $O_3$ formation, consistent with the earlier mentioned observation of ozone. Finally, also the hydrogenation of ozone ($O_3$+H) was studied by Mokrane et al. [089] and Romanzin et al. [094], showing $H_2O$ and $H_2O_2$ formation (as well as their deuterated equivalents). It was found that this reaction channel interacts with the O and $O_2$ hydrogenation reaction scheme. Also this reaction scheme is shown in the figure. At the same time, the O+H/O+D channel was extensively studied by a number of independent groups [085,091,095]. In none of these experiments the impact of $H_2$ was found to result in reaction products at a detectable level [093]. In a very recent study [150], it was also shown that the reaction O+$H_2$ does not largely affect processes in the ISM, which is an important finding, given the large abundances of molecular hydrogen in space. However, Oba et al. [097] showed that reaction OH+$H_2$ proceeds through quantum tunnelling. Other recent studies [100,150] used continuous-time random-walk Monte Carlo simulations to disentangle the different processes that are shown in Figure 4. Simulations are here used to reproduce the experimental findings for two different temperatures – 15 and 25 K – and the two aforementioned different deposition methods. This study confirms that radical diffusion plays an important role in the full network and it shows that the key pathways for the molecular oxygen channel are determined by the $O_2$+H, $HO_2$+H, OH+OH, $H_2O_2$+H and OH+H reactions. The relatively high hydrogen peroxide abundance observed in the experiments is explained by a slow destruction mechanism. The presence of an activation barrier in this reaction was recently confirmed by Oba et al. [102]. Hydrogen peroxide hydrogenation and deuteration rates were compared at 10-30 K and a slower rate was observed for D-atom addition over H-atom addition. This is consistent with the presence of an activation barrier in the reaction that proceeds through quantum tunnelling.

A systematic comparison of the D/H ratios in Earth's oceans, gas-phase cometary water, and gaseous water in proto-planetary disks provides strong evidence that water was delivered to Earth by comets, found in the Oort cloud and composed of the most pristine pre-solar ices. However, a link between water on Earth and water formed on interstellar ice grains cannot be easily established, because (a) unambiguous observational data of deuterated interstellar ices are still lacking, and (b) it is not yet clear to what extent gas-phase deuteration ratios in star-forming regions actually reflect the deuteration ratios in the icy molecular reservoirs from which the gas-molecules originate. In the past few years, there is a growing number of laboratory based studies devoted to the relative comparison between hydrogenation and deuteration rates for selected reaction routes presented in Figure 4 [097,102,155,156].

Meanwhile, also from an astronomical point of view much progress has been made. The WISH key programme of the space telescope Herschel has recently made observations of gas phase water and species related to its chemistry in prestellar cores and young stellar objects at different evolutionary stages. Hot gas containing O(I), $O_2$, cold $H_2O$, $HO_2$, and $H_2O_2$, as well as OH, $OH^+$, $H_2O^+$, and $H_3O^+$ have been identified in the interstellar medium. The identification of these molecules is consistent with the solid state network that has been derived in the laboratory experiments. The astronomical results are summarized in a recent review and linked to the existing laboratory data [157].

### 3.2 CO+H

Solid methanol can act as a starting point for the formation of more complex species when irradiated with UV photons or cosmic rays [039,063]. This finding and its high abundance in the ISM make methanol one of the key molecules in astrochemical reaction schemes. Since gas-phase reaction routes cannot explain the observed methanol abundances, it has to form in a different way [158,159]. Charnley et al. [160] proposed that methanol forms in the solid phase by the hydrogenation of CO via successive reactions of CO with atomic hydrogen: CO → HCO → $H_2CO$ → $H_2COH$ → $CH_3OH$. This surface reaction scheme (shown on the left-side of Figure 4) has been long under debate, as the first sets of laboratory data [085,103,105,106] gave conflicting results. In their first studies, Hiraoka et al. [085,103] reported both the detection of $H_2CO$ and $CH_3OH$ upon hydrogenation of CO ice, but a later study [106] confirmed only the formation of formaldehyde, whereas in [103] an efficient formation of methanol ice was presented. In a follow-up study by Fuchs et al. [025], several years later, covering the 12 - 20 K region and able to vary the H-atom flux, it was shown that the results of both Japanese groups were actually not conflicting at all, but that the slightly different temperatures and particularly flux settings prohibited $CH_3OH$ detection in the Hiraoka experiment.

The first (CO+H) and third ($H_2CO$+H) reaction steps in this H-atom addition scheme possess barriers that have to be overcome. The efficiency of the process therefore depends on the barrier height and how much excess energy is available to cross over. Figure 5 shows the time evolution of the surface abundance (in molecules cm$^{-2}$) of CO, $H_2CO$ and $CH_3OH$ during H-atom bombardment of CO ice with an H-atom flux of $5 \times 10^{13}$ cm$^{-2}$ s$^{-1}$ at different temperatures (**a** 12.0 K, **b** 13.5 K, **c** 15.0 K, and **d** 16.5 K). The amount of CO decreases as the abundance of $H_2CO$ increases for four different temperatures. After bombardment with $1 \times 10^{17}$ H atoms cm$^{-2}$, the formation of methanol kicks-off at the expense of the growth of the $H_2CO$ abundance. Similar trends of abundance evolution as a function of fluence are reported in [111]. Based on the experimental results, in [025], it was concluded that the penetration depth of H atoms in a pure CO ice is at most 4 monolayers (ML). Moreover, for temperatures higher than 15.0 K, a clear drop in the production rate of methanol was observed. This is probably due to two effects: (*a*) desorption of H atoms that becomes important, (*b*) and the reduced sticking of H atoms because of the low H₂ surface abundance at higher temperatures. Both effects cause the H surface abundance to drop substantially at those temperatures and therefore reduce the probability of hydrogenation reactions occurring in the laboratory.

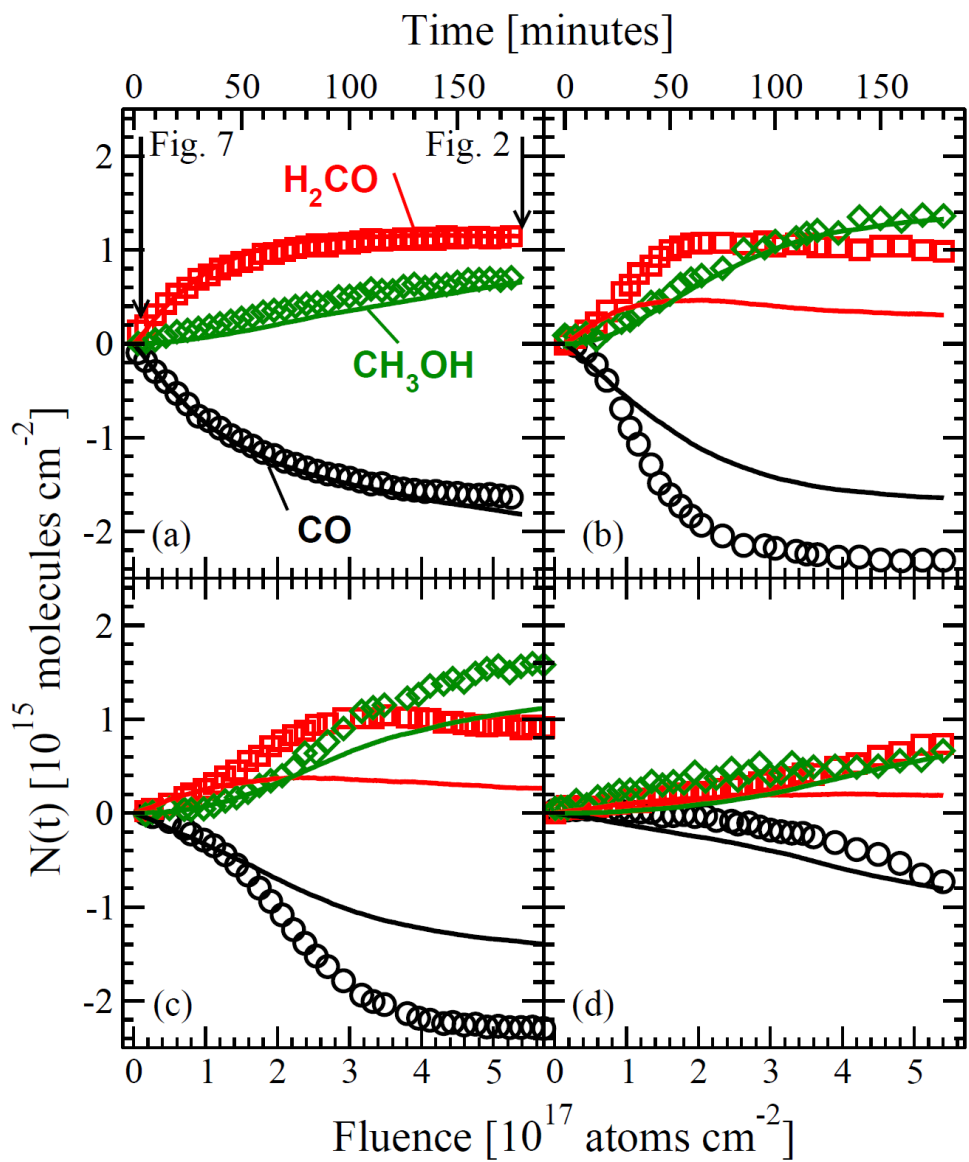

**FIGURE 5** – *Time evolution of the surface abundance (molecules cm⁻²) of CO, H₂CO and CH₃OH during H-atom bombardment of CO ice with an H-atom flux of 5 x 10¹³ cm⁻² s⁻¹ at surface temperatures of 12.0 K (a), 13.5 K (b), 15.0 K (c), and 16.5 K (d). Experimental data (symbols) and Monte Carlo simulation results (solid lines) are shown as well. Figure taken from [025] (reproduced with permission © ESO).*

An interpretation of these measurements in terms of the actual processes taking place – deposition of H atoms, diffusion and then reaction – is only possible using a detailed model. Fuchs et al. used a microscopic kinetic Monte Carlo algorithm to fit effective reaction barriers for CO+H and $H_2CO$+H to a large set of $H_2CO$ and $CH_3OH$ production and CO loss data as function of temperature (see solid lines in Figure 5) and H-atom fluence, yielding typical values in the range of 400 to 500 K. The production rate of $H_2CO$ was found to decrease, and the H-atom penetration depth into the ice to increase with temperature. A similar effect was found for methanol.

Temperature-dependent values are important – within their experimental accuracies – for inclusion in astrochemical models that predict astronomical column densities. As indicated earlier, in such a case the production ratio ($H_2CO$/$CH_3OH$) for the interstellar simulations is a more accurate parameter to use and indeed found to be in closer agreement with observational limits than a direct comparison of the absolute values.

Deuteration experiments were also performed on CO ices [112], which resulted in both deuterated formaldehyde and methanol. It was found that deuteration rate of CO is 12.5 times lower than the rate of CO hydrogenation at the same experimental conditions. A slower rate of CO deuteration in comparison to H-atom addition was also confirmed by Fedoseev et al. [130]. In addition to CO+D experiments, also deuteration studies of $H_2CO$ and $CH_3OH$ ices have been reported [110,113,161]. Figure 1 of Hidaka et al. [113] presents the complete surface reaction network as obtained when CO is exposed to H and D atoms. Here, the reaction routes of $H_2CO$+D and $D_2CO$+H on ASW at 10 - 20 K are studied by IR spectroscopy. For D+$H_2CO$, $H_2CO$ was converted to HDCO and $D_2CO$ by the H–D substitution reactions, and $CD_3OD$ was slightly formed by D atom addition to $D_2CO$. Doubly and triply deuterated methanol, $CH_2DOD$ and $CHD_2OD$, were not observed in the IR spectra, indicating that D atom addition to $H_2CO$ and HDCO is much slower than the H–D substitution reactions. Thus, the H–D substitution reactions in formaldehyde ($H_2CO \rightarrow$ HDCO $\rightarrow D_2CO$) are the dominant reaction route. For H+$D_2CO$, D–H substitution reactions ($D_2CO \rightarrow$ HDCO $\rightarrow H_2CO$) and subsequent $CH_3OH$ formation by H atoms addition to $H_2CO$ proceed efficiently. Doubly deuterated methanol, $CHD_2OH$, which was formed by successive hydrogenation of $D_2CO$, was observed at a significant level. In contrast to D+$H_2CO$, methanol formation is competitive to the D–H substitution reaction. Therefore, the dominant routes are found to be $D_2CO \rightarrow$ HDCO $\rightarrow H_2CO \rightarrow CH_3OH$ and $D_2CO \rightarrow CHD_2OH$. Figure 8 in Hidaka et al. resumes the main routes for deuterated formaldehyde and methanol formation on a dust grain as found from the experiments. The general conclusion here is that although D-atom additions to CO/$H_2CO$ proceed slower than H-atom additions to CO/$D_2CO$, D-atom induced substitution reactions in $H_2CO$ and $CH_3OH$ take place faster than the corresponding H-atom induced substitutions in $D_2CO$ and $CD_3OD$.

The link between solid CO and $CH_3OH$ was also confirmed spectroscopically. It has been a long standing goal to explain the disappearance of the 2152 cm$^{-1}$ CO ice band in astronomical spectra that showed up in all laboratory spectra recording CO in water ice. A study linking bandwidths, peak positions and (dis)appearance of the 2152 cm$^{-1}$ band proved that the astronomical spectra are fully consistent with CO ice intimately mixed with $CH_3OH$, in agreement with a common chemical history as described above [016,162].

### 3.3 CO+OH vs CO+O

The two previous paragraphs describe hydrogenation reactions starting from nearly pure ices. This is generally not the case in space, where ices consist of layered and largely mixed ices. So what happens when a binary ice mixture comprising CO and $O_2$ is hydrogenated? Is the $H_2CO/CH_3OH$ or $H_2O_2/H_2O$ channel dominating? The RAIRS spectrum in Figure 6 shows that for the studied conditions not only the hydrogenation products of pure CO and $O_2$ ice are found, but in addition also $CO_2$ is clearly observed. The appearance of this new molecule is shown in the middle trace of Figure 4 and follows the reaction of CO and OH (formed in the $O_2$ hydrogenation scheme) with subsequent dissociation of the resulting HO-CO complex. Again, also a binary ice is not representative for a 'real' interstellar ice, but this work shows that even in a simple mixed ice the reaction dynamics become quite complicated. The laboratory results also show a correlation between the formation of $CO_2$ and $H_2O$ [163] which is consistent with the astronomical observation of solid $CO_2$ intimately mixed in water-rich environments [164]. Moreover, it is found that the contemporary presence of CO and $O_2$ in the ice influences the final product yields of $H_2CO$, $CH_3OH$, $H_2O_2$ and $H_2O$. The formation rate of the CO hydrogenation reaction products is less affected by the presence of $O_2$ than the $O_2$ hydrogenation reaction products are affected by the presence of CO. This can be explained by the lower penetration depth of H atoms in CO ice and by the formation of $CO_2$ as an additional product, since OH radicals, formed through the $O_2$ channel, are used to form also $CO_2$ instead of $H_2O$ and $H_2O_2$.

The CO+OH channel is one possible pathway to make $CO_2$ in the solid state. $CO_2$ can be formed for instance also along CO+O and HCO+O reaction routes. The latter reaction is experimentally challenging to investigate in the solid phase because other $CO_2$ formation reaction routes will compete. Moreover, when the H/O ratio is in favor of H atoms, the hydrogenation of CO ice will convert most of the HCO in formaldehyde and methanol. Therefore, this route is still to be investigated experimentally, but the reaction CO+O has been extensively studied by different groups [115,116,122].

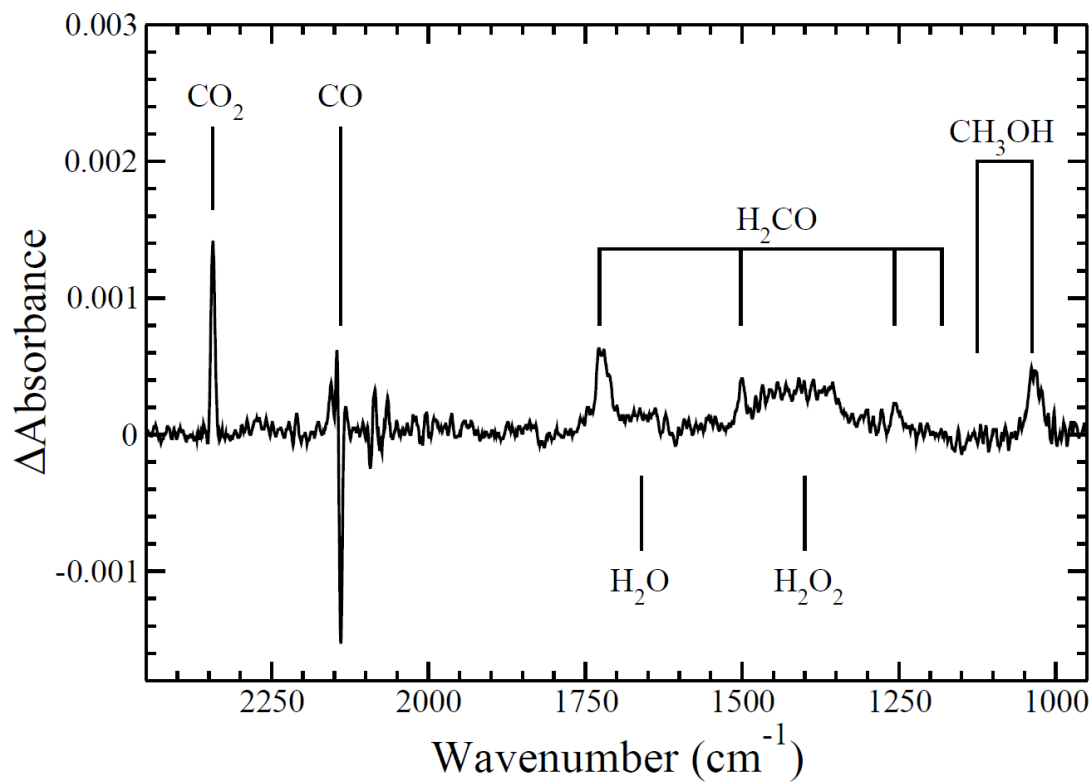

**FIGURE 6** – *RAIR spectrum of a CO:O₂=1:4 ice mixture at 15 K after a H-atom fluence of 1.3 x 10¹⁷ atoms cm⁻². Figure taken from Ioppolo et al. (Fig. 2) [120].*

TPD experiments have been performed to study the CO+O route by using thermal O atoms below 160 K [115]) and energetic O atoms [116]. In the first case, this reaction was found to proceed only in water pores under a water ice cap and upon heating, while in the second case the energetic O atoms allowed the reaction to proceed. More recently, Raut and Baragiola [122] showed by means of IR spectroscopy and microgravimetry that $CO_2$ forms in small quantities during codeposition of CO and cold (non-energetic) O and $O_2$ into thin films at 20 K. The reason for the low $CO_2$ yield is that O atoms react preferentially with O to form $O_2$, and with $O_2$ to form $O_3$. The latter experimental results indicate that this surface reaction has a barrier (~2000 K in the gas phase [165]). Moreover, according to Goumans and Anderson [166] the onset of tunneling is at a too low temperature for the reaction to significantly contribute to the formation of solid $CO_2$ under interstellar conditions. Therefore, CO+O is likely not an efficient pathway to $CO_2$ formation unless energetic processing is involved.

Very recently, Ioppolo et al. [125] directly compared the efficiency of two different carbon dioxide surface formation channels (CO+O and CO+OH) under the same experimental conditions using SURFRESIDE$^2$. The use of a double beam line system is essential here and to interpret results from the simultaneous hydrogenation and oxygenation of solid CO, it is necessary to first distinguish the single contributions of the different reaction channels, i.e., O+H, CO+H, and CO+O. This also has been the topic of a theoretical study [167]

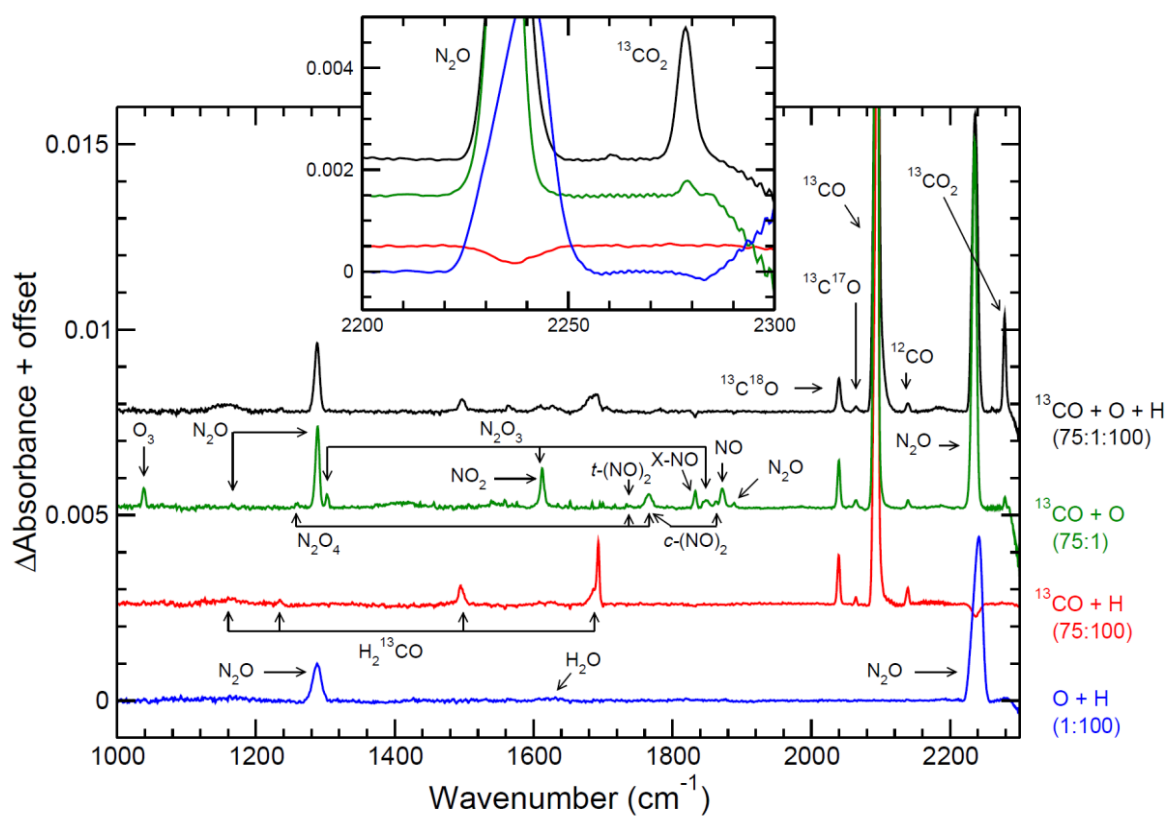

**FIGURE 7** – *RAIR co-deposition spectra of $^{13}CO$, O- and H-atoms on a 13.5 K gold substrate (top-spectrum in main panel, $^{13}CO+O+H$); $^{13}CO$ and O-atoms (second spectrum from the top, $^{13}CO+O$); $^{13}CO$ and H-atoms (third spectrum from the top, $^{13}CO+H$); and O-*

*and H-atoms (bottom-spectrum, O+H). The inset figure shows the $^{13}CO_2$ stretching mode region for all four experiments. (Reproduced with permission from [125] - Copyright [2013], AIP Publishing LLC).*

Figure 7 shows the IR spectrum of $^{13}CO$ co-deposited with oxygen and hydrogen atoms on a 13.5 K gold substrate ($^{13}CO+O+H$). This spectrum is compared to co-deposition spectra of $^{13}CO$ and oxygen atoms ($^{13}CO+O$); $^{13}CO$ and hydrogen atoms ($^{13}CO+H$); and oxygen atoms and hydrogen atoms (O+H). In all four experiments, oxygen atoms are produced by the MWAS, while hydrogen atoms are formed by the HABS. The $^{13}CO$ deposition rate (0.0075 ML s$^{-1}$) as well as O- and H-atom fluxes (~$10^{11}$ and ~$10^{13}$ atoms cm$^{-2}$ s$^{-1}$, respectively) are kept the same in all the experiments and the spectra shown in Figure 7 are all acquired after 45 minutes of co-deposition.

Under the same experimental conditions $CO_2$ is found to form more efficiently through the reaction CO+OH than CO + O, indicating that the first reaction has a lower activation barrier than reaction CO+O. These results are not only in good agreement with other experimental results (CO+O, [122]; CO+OH, [118,120,121]), but also with the most recent astrochemical models and observations [166,168]. In the latter study, Chang and Herbst studied, among others, the surface reaction CO+O+H, by means of a unified microscopic-macroscopic Monte Carlo simulation of gas-grain chemistry in cold interstellar clouds in which both the gas-phase and the grain-surface chemistry are simulated by a stochastic technique. In their model, solid $CO_2$ is produced mainly by the reaction (CO+OH), which occurs by a so called "chain reaction mechanism", in which an H atom first combines with an O atom lying above a CO molecule, so that the OH does not need to undergo horizontal diffusion to react with CO. This scenario is not far from the experimental conditions described in [125], where O and H atoms meet to form OH radicals that then further react with neighboring CO molecules to form $CO_2$. Chang and Herbst conclude that the solid CO formed in early cold cloud stages via accretion and surface reactions is mainly converted into $CO_2$ through reaction (CO+OH). This makes the latter reaction to be most likely the main non-energetic $CO_2$ formation route under early cold cloud conditions, where H atoms are considerably more abundant than O atoms [169]. Chang and Herbst [168] also suggested that the conversion of CO into $CO_2$ becomes inefficient at later times, where, for the low-mass YSO case, there can be a high abundance of almost pure CO, with some conversion to formaldehyde and methanol. Under these conditions, solid $CO_2$ can still be formed via energetic processing (UV and cosmic ray irradiation; e.g., [170], and references therein.

Although dense molecular clouds are shielded from UV radiation to a great extent by dust particles, cosmic rays can still penetrate these regions and, therefore, they can efficiently process the ices (e.g., [171-173]). As discussed in section 2, energetic processing becomes even more important at later stages of star forming regions, when ices are exposed to the irradiation of a new born star. In dense molecular clouds, cosmic rays generate UV photons as well as fast ions and electrons that can indeed release their energy to the target material. Owing to the interaction with fast ions molecular bonds are broken and, on timescales of tens of picoseconds, the molecular fragments recombine giving rise to a rearrangement of the chemical structure that leads to forming new molecular species. In the case of UV photolysis, the energy is released to the target material through a single photo-dissociation or photo-excitation event (see section 2). Therefore, energetic processing most likely contributes to the total $CO_2$ abundance observed in polar and apolar interstellar ices. For instance, (a) photo-dissociation of $H_2O$ ice is a possible mechanism to form available OH radicals that can subsequently react with CO molecules to form solid $CO_2$ through reaction CO+OH; (b) $CO_2$ can be formed through reaction CO+O in apolar ices when the reaction is induced by energetic input and the formed O atom is in an electronically excited state.

Outside the main topic of this review, it should be noted that over the past two decades, several laboratory experiments from different groups have shown that $CO_2$ is efficiently formed after energetic processing of pure CO ice and ice mixtures containing CO and $H_2O$ (e.g., [060,174-178]). Furthermore, $CO_2$ ice is also found upon irradiation of carbon grains covered by a water cap or an oxygen layer [179-182]. Ioppolo et al. [183] and Garozzo et al. [184] quantitatively studied the formation of $CO_2$ ice upon ion irradiation of interstellar relevant ice mixtures containing C- and O-bearing species at low (12-15 K) and high (40-60 K) temperatures, respectively.

### 3.4 Formation of carbon bearing acids and ethanol

In the past few years, several other surface reaction schemes that lead towards a higher molecular complexity have been investigated under laboratory conditions. One of those molecules is formic acid (HCOOH), the smallest organic acid that has been observed towards high- and low-mass star-forming regions and quiescent clouds in the gas phase and likely also in the solid phase (e.g., [185-188]).

Since dissociative recombination of protonated formic acid has been proven not to be an efficient gas-phase channel leading to HCOOH formation, HCOOH is likely to form through surface reactions on dust grains [189]. Although a number of different surface channels has been proposed in the past ($HCO_2$+H, [084,190] and HCO+OH, [191]), the formation pathway to solid HCOOH was not experimentally confirmed until 2011. In Ioppolo et al. [143] the first laboratory evidence is presented for the efficient HCOOH formation during co-deposition experiments of H atoms and $CO:O_2$ mixtures with 4:1, 1:1 and 1:4 ratios. During co-deposition, spectral changes in the ice were monitored by means of a Fourier transform infrared spectrometer using RAIRS. After co-deposition a TPD experiment was performed and gas-phase molecules were detected by a QMS. Formation of HCOOH was observed at low temperatures mainly through hydrogenation of the HO–CO complex, while reactions with the HCO radical as intermediate were found to be inefficient. The HO–CO complex channel, which was previously not considered as an important HCOOH formation route in astrochemical models, can actually explain the presence of HCOOH in dense cold clouds, at the beginning of the warm-up phase of a protostar, and, therefore, is likely to be astrochemically relevant. Moreover, the HO–CO+H reaction channel links HCOOH to the surface formation of $CO_2$ and $H_2O$ with a similar formation branching ratio for the three final products.

As was shown by Oba et al. [144], solid $H_2CO_3$ can also be formed from the same HO–CO complex through the reaction HO–CO+OH. This reaction leads to the formation of $CO_2+H_2O$ and $H_2O_2+CO$. The fact that in [143] solid $H_2CO_3$ was not detected, and that in [144] no HCOOH was detected, is probably due to a different atom/radical flux composition: in the latter experiment OH radicals are produced in a plasma of $H_2O$, while in the former one OH radicals are formed in the ice through surface reactions of $O_2+H$. Therefore, in the first case there is likely an overabundance of OH radicals on the surface, while in the second case H is the dominant species, thus HO–CO+H is more likely to occur.

Bisschop et al. [145] investigated the hydrogenation of solid $CO_2$, HCOOH and $CH_3CHO$ under interstellar relevant conditions. RAIRS and TPD were used to analyze the results. $CO_2$ and HCOOH were found to not react with H-atoms at a detectable level. Therefore, a solid state formation of HCOOH from $CO_2$ and $CH_2(OH)_2$ is likely inefficient in interstellar ices. Hydrogenation of $CH_3CHO$ leads to ~20% of $C_2H_5OH$, showing for the first time that a thermal hydrogenation reaction can be responsible for the ethanol abundances detected in dense interstellar clouds.

## 3.5     Formation of C-C bond bearing complex organic molecules by hydrogenation of CO molecules

A clear focus of the past few years has been on the astronomical detection of amino acids, specifically the simplest amino acid, glycine. Despite theoretical studies and laboratory based work that show that glycine as well as several other amino-acids should form in space [033,127,192] unambiguous detections – even though claimed regularly - are still lacking [193]. The search for two other classes of prebiotic compounds – aldoses (polyhydroxy aldehydes) and polyols – has been more successful. Aldoses are compounds with chemical formula $(CH_2O)_n$ containing one aldehyde (-CHO) group. Well-known members of this series are the simple sugars (monosaccharides) such as glycose, ribose and erythrose. The simplest representative of this class – glycoladehyde ($HC(O)CH_2OH$) – has been successfully detected in the ISM [194]. The best-known representative of the polyols series is glycerine – a basic compound of fats. Glycerine is a triol and has not been detected in space so far, but ethylen glycol ($H_2C(OH)CH_2OH$), a diol, has been observed [195,196]. Fedoseev et al. [130] showed the first laboratory evidence of the formation of two **larger** molecules of astrobiological importance – glycolaldehyde ($HC(O)CH_2OH$) and ethylene glycol ($H_2C(OH)CH_2OH$) – by surface hydrogenation of CO molecules, *i.e* along a non-energetic process merging a backbone from simple species; the key step in this process is believed to be the recombination of two HCO radicals followed by the formation of a C-C bond and follows theoretical work on glycolaldehyde [197]. This is an important experimental finding, as so far hydrogenation reactions were mainly shown to be effective in the formation of smaller species (e.g., ammonia from N+H) with $CH_3OH$ (CO+H) as the largest species systematically studied so far by more than one independent group.

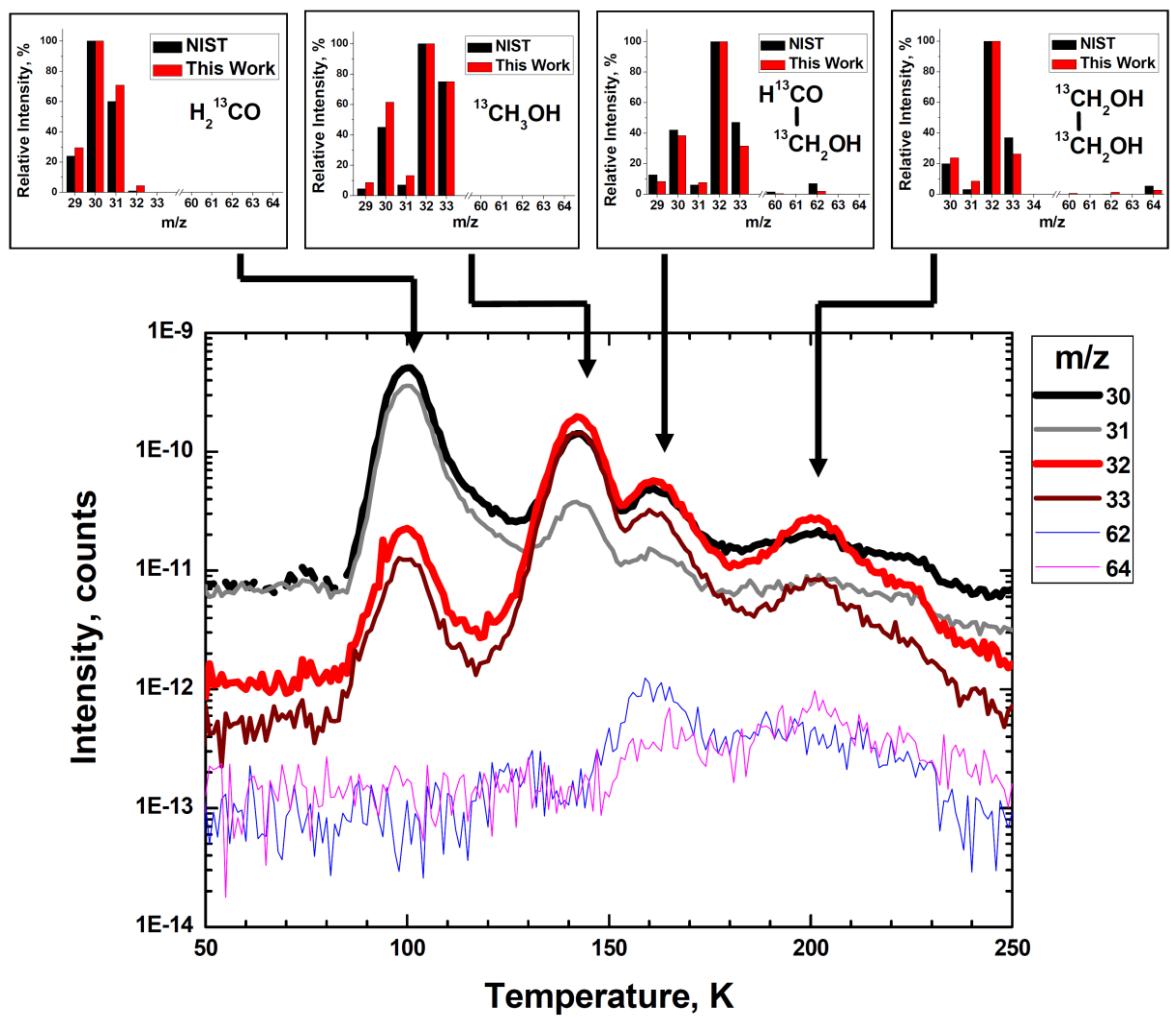

**FIGURE 8 –** *TPD spectra obtained after a codeposition experiment of $^{13}CO+H$ (1:25) at 13 K for 360 minutes for the indicated m/z numbers. Insets on top of the figure show the comparison between fragmentation patterns of the desorbing species detected in this experiment upon 60 eV electron impact with those available from literature. Figure taken from Fedoseev et al. (Fig. 3) [130].*

In Figure 8, a typical example of a TPD quadrupole mass spectrum is presented for a $^{13}$CO + H codeposition experiment at 13 K. Four peaks are found in the TPD QMS spectra. The insets in the top-side of the figure compare the fragmentation patterns of desorbing species recorded in this experiment with values obtained by extrapolating literature results. The experimental and literature values are very close and fully consistent with the finding that glycolaldehyde (H$^{13}$C(O)$^{13}$CH$_2$OH) and ethylen glycol (H$_2$$^{13}$C(OH)$^{13}$CH$_2$OH) form along with the formation of H$_2$$^{13}$CO and $^{13}$CH$_3$OH upon surface hydrogenation of CO molecules. $^{13}$CO experiments are used to provide additional proof for glycolaldehyde and ethylene glycol formation because, with $^{13}$CO as precursor, dissociative ionization products containing only one carbon atom will be shifted by one m/z number, while fragments containing two carbon atoms will shift by two. This is a characteristic property for C-C bond formation.

These experiments aim at simulating the CO freeze-out stage in interstellar dark cloud regions, well before thermal and energetic processing become dominant. Therefore, Fedoseev et al. [130] show that formation of both glycolaldehyde and ethylene glycol may take place already in the prestellar stage well before energetic processing of the ice by the newly formed protostar will take place. Moreover, the authors found that the correlation between the abundances of HC(O)CH$_2$OH and H$_2$C(OH)CH$_2$O with those of formaldehyde and methanol is as expected typically in ratios of the order of a few percent.

At this place, it should be stressed that the solid state formation of complex organic material has been a very active research area over many decades and particularly the VUV irradiation or cosmic ray bombardment of astrophysically relevant ice mixtures has been taken as a proof that biologically relevant species can form in the solid state. As stated in section 1.2, such studies follow both top-down and bottom-up approaches, and the formation of specifcally glycolaldehyde and ethylene glycol also has been observed upon energetic processing of methanol containing ices [039,053,063,198,199]. Other pathways for $HC(O)CH_2OH$ and $H_2C(OH)CH_2O$ formation have been predicted by theory [200-202]. The relevance of the atom addition channel presented here is particularly interesting as it takes place along a non-energetic process and consequently may be relevant for chemical pathways in dark clouds.

## 3.6    Surface chemistry of nitrogen bearing molecules

Ammonia ($NH_3$) is one of the few nitrogen-bearing species that have been observed in interstellar ices toward young stellar objects (YSOs) and quiescent molecular clouds. Although the nitrogen chemistry of the interstellar medium (ISM) has the potential to reveal the link between the formation of simple species and prebiotics, like amino acids, this is still poorly investigated in laboratories and therefore not well understood. The sequential hydrogenation of N atoms was first tested at cryogenic temperatures by Hiraoka et al. [138] who performed a TPD experiment after the hydrogenation of N atoms trapped in a matrix of solid $N_2$. Hidaka et al. [139] confirmed the formation of ammonia in a solid $N_2$ matrix at low temperatures. In this study, however, the unambiguous detection of $NH_3$ was made only after annealing the ice to 40 K in order to desorb the $N_2$ matrix. Very recently, Fedoseev et al. [140,141] studied the surface formation of $NH_3$ through sequential hydrogenation of N atoms at low temperatures and in interstellar relevant ice analogues. They found the reactions to proceed through a Langmuir-Hinshelwood mechanism, to be fast and likely barrierless, thus constraining the previous findings of the different groups. The surface $NH_3$ formation mechanism is therefore relevant to dense cloud conditions, where ices are shielded from radiation and a polar (water-rich) ice is formed on top of the bare grains. The reactivity of the intermediate NH and $NH_2$ radicals embedded in more complex ices is also experimentally investigated in this work, and the formation of HNCO in a CO ice is shown. The latter experiments are somehow closer to the CO freeze-out stage conditions observed in space, when CO rapidly accretes on top of the previously formed $H_2O$–rich ice mantle and forms a layer of CO-rich non-polar ice. When ammonia and isocyanic acid are present in the ice and are heated to higher temperatures, they react to form $NH_4^+$ and $OCN^-$ [75,77]. The formation of these ions is indeed commonly associated with a later stage of molecular cloud evolution, when thermal processing of the ice by a newly formed protostar becomes important. These laboratory results [140,141] are in good agreement with observations [008] and have potential inplications in astrobiology as HNCO and $OCN^-$ are precursors of simple amino acids and peptide fragments. In their studies, Fedoseev et al. [141] further investigated the deuterium enrichment of the formed ammonia ice by co-depositing H, D, and N atoms on the cold substrate. This suggested that the deuterium enrichment of the ice is likely due to the higher binding energy of D atoms to the ice surface compared to H atoms.

Similarly to the C-H bonds of $H_2CO$ and $CH_3OH$, an N-H bond may be subjected to D-atom induced exchange. Although $NH_3$ itself does not participate in H-D substitutions [112,141] at cryogenic temperatures, Oba and coworkers [203] have recently investigated experimentally the H–D and D–H substitutions of solid methylamine isotopologues through surface reactions at 10 K. They showed that both H–D and D–H substitution occur faster in the methyl group of methylamines than in the amino group. One possible explanation for this result is that a series of H(D) abstraction–D(H) addition reactions is the main process for H–D (D–H) substitution in the methyl group. In addition to the substitution–addition process, the formation of an intermediate species, like $CH_3NH_2D$, followed by the reaction with D(H) atoms may also be effective for the H–D (D–H) substitution in amino groups. Based on the effective rate constants for H–D and D–H substitution reactions experimentally obtained in this study, Oba et al. [203] predicted that singly deuterated methylamine CH$_2$DNH$_2$ is the most abundant isotopologue formed by surface reactions within the typical lifetime of a molecular cloud.

Hydroxylamine (NH$_2$OH) is another of the potential precursors of complex prebiotic species in space (glycine and β-alanine; [192]). In 2012, a detailed experimental study of hydroxylamine formation through nitric oxide (NO) surface hydrogenation was performed with particular care to the submonolayer regime on interstellar relevant substrates (crystalline H$_2$O and amorphous silicate) and the multilayer regime in interstellar relevant (H$_2$O- and CO-rich) ices [126-128]. NH$_2$OH was found to be efficiently formed through a barrierless reaction mechanism. Figure 9 shows RAIR spectra of solid NO co-deposited on the gold substrate at 15 K with low- and high-hydrogen atom flux (H/NO = 0.2 and H/NO=4, respectively). The non-detection of NH$_2$OH in the gas phase indicates that hydroxylamine is likely directly converted to other species on the ice surface before desorption or, alternatively, efficiently consumed through gas-phase reactions with H$_3^+$ and CH$_5^+$ [192,204].

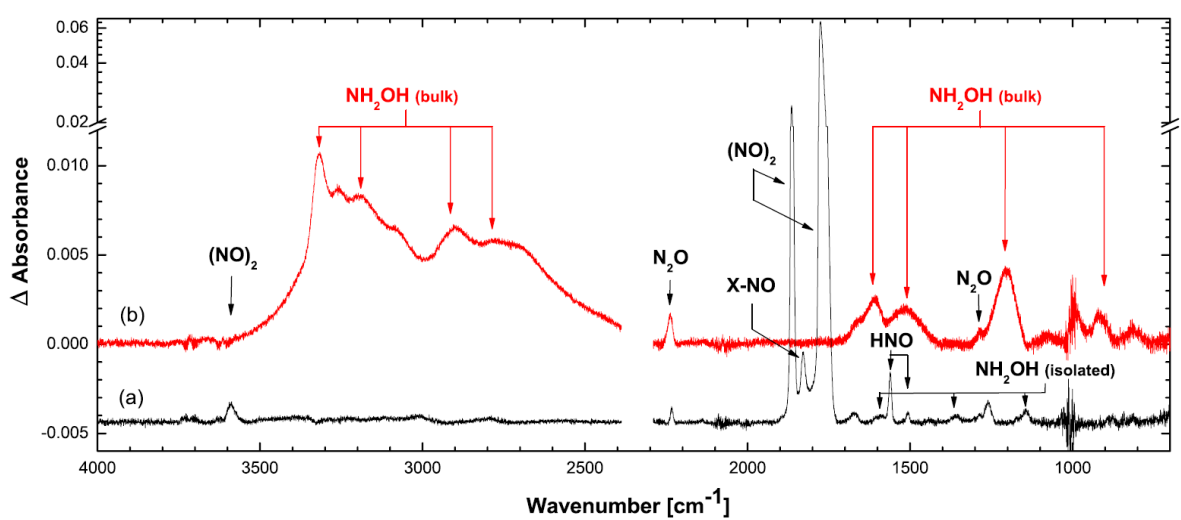

**FIGURE 9** – *RAIR spectra of solid NO deposited on a gold substrate at 15 K upon H-atom exposure: (a) co-deposition spectrum of NO and H atoms after 60 minutes and with a 'low' H-atom flux of 7 × 10$^{12}$ atoms cm$^{−2}$ s$^{−1}$ (H/NO = 0.2); (b) co-deposition spectrum of NO and*

*H atoms after 120 minutes and with a 'high' H-atom flux of 3 × 10$^{13}$ atoms cm$^{-2}$ s$^{-1}$ (H/NO = 4). Figure taken from [126] (reproduced with permission IOP).*

Related to the hydrogenation studies of NO is a large set of experiments in which H-, O-, and N-atom addition reactions have been investigated [131,133,134]. These experiments extended the hydroxylamine reaction network including reactions NO+H/N/O/O$_2$/O$_3$ and NO$_2$+H/N/O that lead to the formation of water, hydroxylamine, and nitrogen oxides (NO, NO$_2$, N$_2$O, N$_2$O$_3$, N$_2$O$_4$) on cold (10-15 K) surfaces (see Figure 10). Finally, He et al. [129] showed the laboratory formation of hydroxylamine on dust grain analogues via ammonia oxidation, adding a new channel to the formation of a glycine precursor.

Due to its link to the formation of prebiotic species, solid state nitrogen chemistry will surely be the subject of several future astrochemical studies, combining laboratory data with observations and models, with the goal to "unlock the chemistry of the heavens".

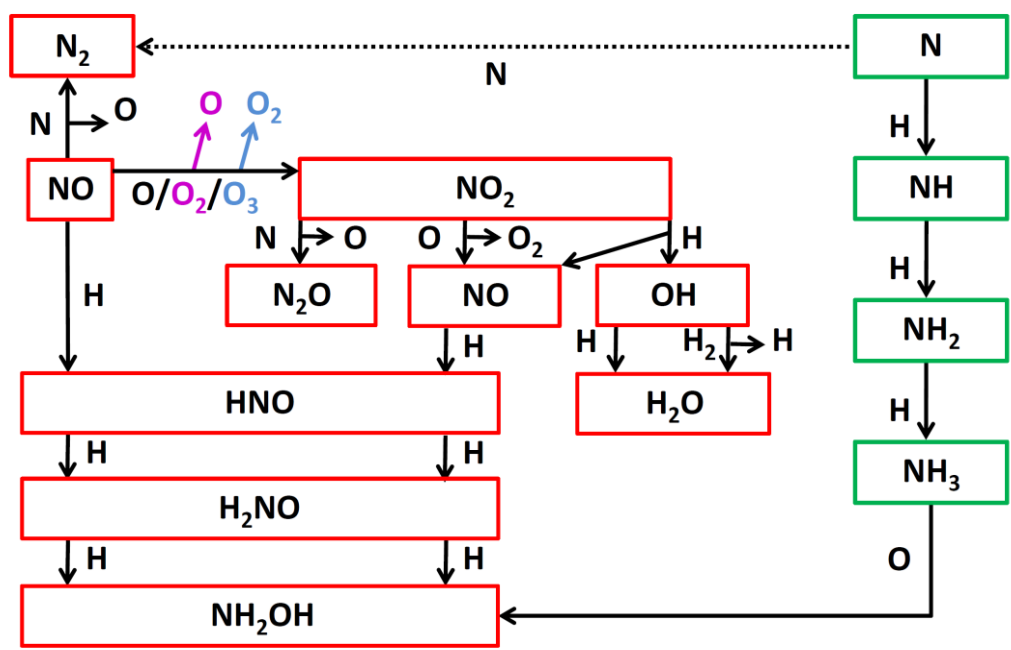

**FIGURE 10 –** *The complete nitrogen chemistry network, summarizing the findings from [126-134].*

### 4. Conclusions

The take home message of this review article is that atom addition reactions in interstellar ice analogues, as investigated in the laboratory, provide solid state reaction pathways that are of relevance to explain the astronomical observation of a large number of molecules in the ISM. Dedicated experiments allow investigating isolated reaction schemes in full dependence of relevant settings and sensitive diagnostic tools make it possible to extend qualitative trends to quantitative conclusions. Physical and chemical parameters are derived that are needed as input in astrochemical models and that allow to bridge conclusions from a laboratory setting to interstellar space. The level of complexity is quite overwhelming, even for relatively simple ice mixtures. In Figure 11 this is illustrated by connecting all reaction schemes discussed here. The arrows show possible reaction pathways, but one has to realize that their efficiency depends on temperature (for example), and on the presence of specific precursors that will depend on the evolutionary stage of the process. Other processes not studied here are likely to interact as well. Nevertheless, work as described here is necessary to extend the existing databases with solid state parameters, urgently needed, to fully understand how gas/grain and grain/gas interactions influence the evolution of the Molecular Universe.

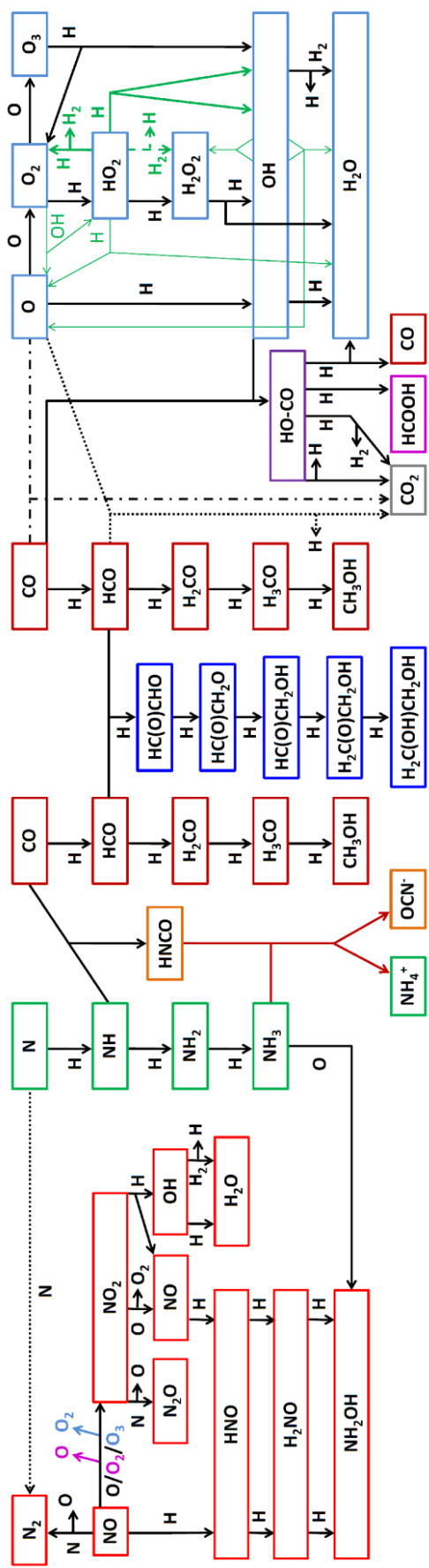

**FIGURE 11 −** *A summary of the atom addition reactions discussed in this review. The arrows indicate possible pathways, but it should be noted that their efficiency and influence on molecule abundances in space depends on a number of different parameters. Moreover, other (energetic) processes are at play as well. The take home message is that solid state chemistry is complex and that non-energetic ice processing offers a pathways to produce (the building blocks to form) complex molecules.*

**Acknowledgement**

The study of chemical processes in interstellar ice analogues is a key research area in the Sackler Laboratory for Astrophysics at Leiden Observatory. Many PhD students and postdocs as well as visitors have been involved in performing and interpreting experiments on one or more of our five experimental setups fully dedicated to the spectroscopy and dynamics of interstellar ices (www.laboratory-astrophysics.eu). Here we specifically thank the people involved on SURFRESIDE(2) related work: Suzanne Bisschop, Guido Fuchs, Herma Cuppen, Claire Romanzin, Thanja Lamberts, and KoJu Chuang. We have benefitted a lot from discussions with Profs. Ewine van Dishoeck and Xander Tielens. Much of the work described here has been performed within the framework of i) LASSIE, a large FP7-ETN network (GA 238258), ii) Network 2 science of NOVA, the Dutch Research School for Astronomy, iii) the solid state thee of the Dutch Astrochemistry Network, a national NWO program, and iv) a VICI grant of HL. Furthermore, SI acknowledges funding through a Marie Curie Fellowship (FP7-PEOPLE-2011-IOF-300957) and recent support by the Royal Society.